\documentclass[pra,aps,longbibliography,showpacs,groupedaddress,superscriptaddress,floatfix,twocolumn,nofootinbib,toc=flat]{revtex4-1}
\usepackage{float}

\usepackage{amsfonts,amsmath,amssymb,stmaryrd}

\usepackage[utf8]{inputenc}
\usepackage[table]{xcolor}
\usepackage{bbold, bm}
\usepackage{graphicx}
\usepackage{array,multirow}
\usepackage{algorithm}
\usepackage{algpseudocode}
\algrenewcommand\algorithmicrequire{\textbf{Input:}}
\algrenewcommand\algorithmicensure{\textbf{Output:}}
\usepackage{boldline}
\usepackage{tabularx}
\usepackage{tikz}
\usepackage{placeins}
\usepackage{mathrsfs}
\pdfmapline{-dummy LMRoman10-Regular} %avoid disappearing characters in embedded PDFs
\pdfmapline{-dummy LMRoman10-Bold} %avoid disappearing characters in embedded PDFs
\pdfmapline{-dummy LMRoman10-Italic} %avoid disappearing characters in embedded PDFs
\pdfmapline{-dummy LMRoman10-BoldItalic} %avoid disappearing characters in embedded PDFs
\pdfmapline{-dummy LMRoman12-Regular} %avoid disappearing characters in embedded PDFs
\pdfmapline{-dummy LMRoman12-Bold} %avoid disappearing characters in embedded PDFs
\pdfmapline{-dummy LMRoman12-Italic} %avoid disappearing characters in embedded PDFs
\pdfmapline{-dummy LMRoman12-BoldItalic} %avoid disappearing characters in embedded PDFs

\usepackage{ctable}
\usepackage{epsfig}
\usepackage[english, status=final]{fixme}
\fxsetup{marginface=\tiny}
\fxusetheme{color}    

\usepackage[colorlinks=true,linkcolor=blue,urlcolor=blue,citecolor=blue]{hyperref}

\newcommand{\Gj}{\hat{G}_{\bm{j}}}

\newcommand{\bra}[1]{\langle#1|}
\newcommand{\ket}[1]{|#1\rangle}

\renewcommand{\j}{\bm{j}}

\newcommand{\nj}{\hat{n}_{\bm{j}}}

\newcommand{\Ztwo}{$\mathbb{Z}_2$}

\makeatletter
\def\maketitle{
\@author@finish
\title@column\titleblock@produce
\suppressfloats[t]}
\makeatother

\AtBeginDocument{%
    \newwrite\bibnotes
    \def\bibnotesext{Notes.bib}
    \immediate\openout\bibnotes=\jobname\bibnotesext
    \immediate\write\bibnotes{@CONTROL{REVTEX41Control}}
    \immediate\write\bibnotes{@CONTROL{%
    apsrev41Control,author="08",editor="1",pages="1",title="0",year="1"}}
     \if@filesw
     \immediate\write\@auxout{\string\citation{apsrev41Control}}%
    \fi
}%

\begin{document}

\title{Independent e- and m-anyon confinement in the \\ parallel field toric code on non-square lattices}

\author{Simon M. Linsel}
\email{simon.linsel@lmu.de}
\affiliation{Department of Physics and Arnold Sommerfeld Center for Theoretical Physics (ASC), Ludwig-Maximilians-Universit\"at M\"unchen, Theresienstr. 37, M\"unchen D-80333, Germany}
\affiliation{Munich Center for Quantum Science and Technology (MCQST), Schellingstr. 4, D-80799 M\"unchen, Germany}

\author{Lode Pollet}
\affiliation{Department of Physics and Arnold Sommerfeld Center for Theoretical Physics (ASC), Ludwig-Maximilians-Universit\"at M\"unchen, Theresienstr. 37, M\"unchen D-80333, Germany}
\affiliation{Munich Center for Quantum Science and Technology (MCQST), Schellingstr. 4, D-80799 M\"unchen, Germany}

\author{Fabian Grusdt}
\affiliation{Department of Physics and Arnold Sommerfeld Center for Theoretical Physics (ASC), Ludwig-Maximilians-Universit\"at M\"unchen, Theresienstr. 37, M\"unchen D-80333, Germany}
\affiliation{Munich Center for Quantum Science and Technology (MCQST), Schellingstr. 4, D-80799 M\"unchen, Germany}

\date{\today}
\begin{abstract}
Kitaev's toric code has become one of the most studied models in physics and is highly relevant to the fields of both quantum error correction and condensed matter physics. Most notably, it is the simplest known model hosting an extended, deconfined topological bulk phase. To this day, it remains challenging to reliably and robustly probe topological phases, as many state-of-the-art order parameters are sensitive to specific models and even specific parameter regimes. With the emergence of powerful quantum simulators which are approaching the regimes of topological bulk phases, there is a timely need for experimentally accessible order parameters. Here we study the ground state physics of the parallel field toric code on the honeycomb, triangular and cubic lattices using continuous-time quantum Monte Carlo. By extending the concept of experimentally accessible percolation-inspired order parameters (POPs) we show that electric and magnetic anyons are independently confined on the honeycomb and triangular lattices, unlike on the square lattice. Our work manifestly demonstrates that, even in the ground state, we must make a distinction between topological order and (de-)confinement. Moreover, we report multi-critical points in the aforementioned confinement phase diagrams. Finally, we map out the topological phase diagrams on the honeycomb, triangular and cubic lattices and compare the performance of the POPs with other topological order parameters. Our work paves the way for studies of confinement involving dynamical matter and the associated multi-critical points in contemporary quantum simulation platforms for $\mathbb{Z}_2$ lattice gauge theories.
\end{abstract}
\maketitle

%%%%%%%%%%%%%%%%%%%%%%%%
% Introduction
%%%%%%%%%%%%%%%%%%%%%%%%

\section{Introduction}

Quantum models featuring topological order \cite{Wen1989, Wen1990, Wen2007} are one of the main research directions in modern condensed matter physics. They are highly relevant for the study of the integer \cite{Klitzing1980} and fractional \cite{Laughlin1983} quantum Hall effect, quantum spin liquids \cite{Balents2010} and quantum error correction \cite{Kitaev2003}. Topological phases have intriguing features such as degenerate, long-range entangled ground states and point-like excitations (``anyons'') in two-dimensional systems, which fulfill neither fermionic nor bosonic statistics but can instead pick up \textit{any} phase (hence the name) when braiding two anyons \cite{Leinaas1977, Wilczek1982}. As a result, the topological phases remain stable against local perturbations and thus constitute an important class of models for fault-tolerant quantum computing \cite{Kitaev2003, Nayak2008}.

The toric code, originally studied by Kitaev \cite{Kitaev2003}, is generally regarded as the simplest model featuring $\mathbb{Z}_2$ topological order and anyons. Fradkin and Shenker \cite{Fradkin1979} famously studied the extended toric code, i.e. the toric code in a parallel field, on the square lattice. At zero temperature, it features two phases separated by a continuous phase transition: an extended topological deconfined phase for small fields and a trivial confined phase for large fields which encloses a first-order line that ends at a multi-critical point \cite{Fradkin1979, Wu2012} whose universality class is still a topic of debate \cite{Kitaev2003, Vidal2009, Tupitsyn2010, Gazit2018, Somoza2021, Bonati2022_2, Bonati2022, Iqbal2022, Manoj2023, Oppenheim2024}. Other recent studies suggest that the qualitative structure of the phase diagram also holds on the honeycomb, triangular, and cubic lattices \cite{Reiss2019, Kott2024}, yet numerically exact studies have not been reported. 

There is a timely need for experimentally accessible order parameters due to the emergence of quantum simulators enabling snapshot measurements \cite{Zohar2017, Barbiero2019, Schweizer2019_2, Homeier2021, Homeier2022, Mildenberger2025, Halimeh2025} and moving towards the regimes of quantum spin liquids \cite{Semeghini2021}. Topological phases inherently involve non-local entanglement and importantly cannot be probed using local order parameters known from the Ginzburg-Landau paradigm. In the last decades, a plethora of methods have been established to probe topological phases, ranging from topological entanglement entropy \cite{Kitaev2006, Levin2006, Chen2010} to string-loop order operators derived from Wegner-Wilson \cite{Wegner1971, Wilson1974} and 't~Hooft loops \cite{Thooft1978}. A fundamental challenge is that most topology probes are tailored to specific analytical, numerical or experimental frameworks and are not easily accessible to other methods. E.g., the topological entanglement entropy can usually be extracted with the density matrix renormalization group (DMRG) \cite{Schollwoeck2011} or wavefunction-based approaches, however, it requires a gap and is challenging to extract from both quantum Monte Carlo (requiring the replica trick \cite{Melko2010}) and experiments \cite{Ott2024}.

Here we map out the topological and confinement phase diagrams of the extended toric code on the honeycomb, triangular and cubic lattices using a numerically exact state-of-the-art continuous-time quantum Monte Carlo (QMC) algorithm \cite{Wu2012}. We demonstrate that e-anyons and m-anyons are independently confined on the triangular and honeycomb lattices. To this end, we generalize the recently proposed percolation-inspired order parameters (POPs) \cite{Linsel2024, Duennweber2025} -- which are experimentally accessible to snapshot measurements and were shown to capture the phases of the Fradkin-Shenker model -- from e-anyons to m-anyons. The topological phase is identified with the phase where both e- and m-anyons are deconfined \cite{Fradkin1979}, see Fig.~\ref{figAnyons}. We find a set of multi-critical points in the topological and confinement phase diagrams that closely resemble the structure of the Fradkin-Shenker phase diagram. Finally, we compare the POP performance with two other order parameters: a staggered imaginary times observable (SIT) and the Fredenhagen-Marcu loop operator (FM). %As all investigated order parameters are basis-dependent and generally only applicable in certain parameter regimes in a given basis \cite{Linsel2024, Xu2024}, we simulate the extended toric code (\ref{eq:eTC}) in both the $\hat{\tau}^x$- and $\hat{\tau}^z$-bases. 

%owing to the complex nature of topology

%%%%%%%%%%%%%%%%%%%%%%%%
% Main text
%%%%%%%%%%%%%%%%%%%%%%%%

\section{Extended toric code}

We study the extended toric code, described by the Hamiltonian 
\begin{align} \label{eq:eTC}
    \hat{\mathcal{H}} = &- \sum_v \hat{A}_v \; - \sum_{p} \hat{B}_p \nonumber \\
    &- h_x \sum_{l} \hat{\tau}_l^x \; - h_z \sum_{l} \hat{\tau}_l^z,
\end{align}
where Pauli matrices $\hat{\tau}_l^x$ and $\hat{\tau}_l^z$ are defined on the links $l$ of the underlying lattice. The star term $\hat{A}_v = \prod_{l \in v} \hat{\tau}_l^x$ describes the interaction of all links $l$ connected to a vertex $v$ and the magnetic term $\hat{B}_p = \prod_{l \in p} \hat{\tau}_l^z$ describes the interaction of links on an elementary plaquette $p$ of the respective lattice. The terms $\propto h_x, h_z$ are external fields. In this work, we only consider $h_x, h_z \geq 0$ and periodic boundaries.

\begin{figure}[]
\includegraphics[width=0.49\textwidth]{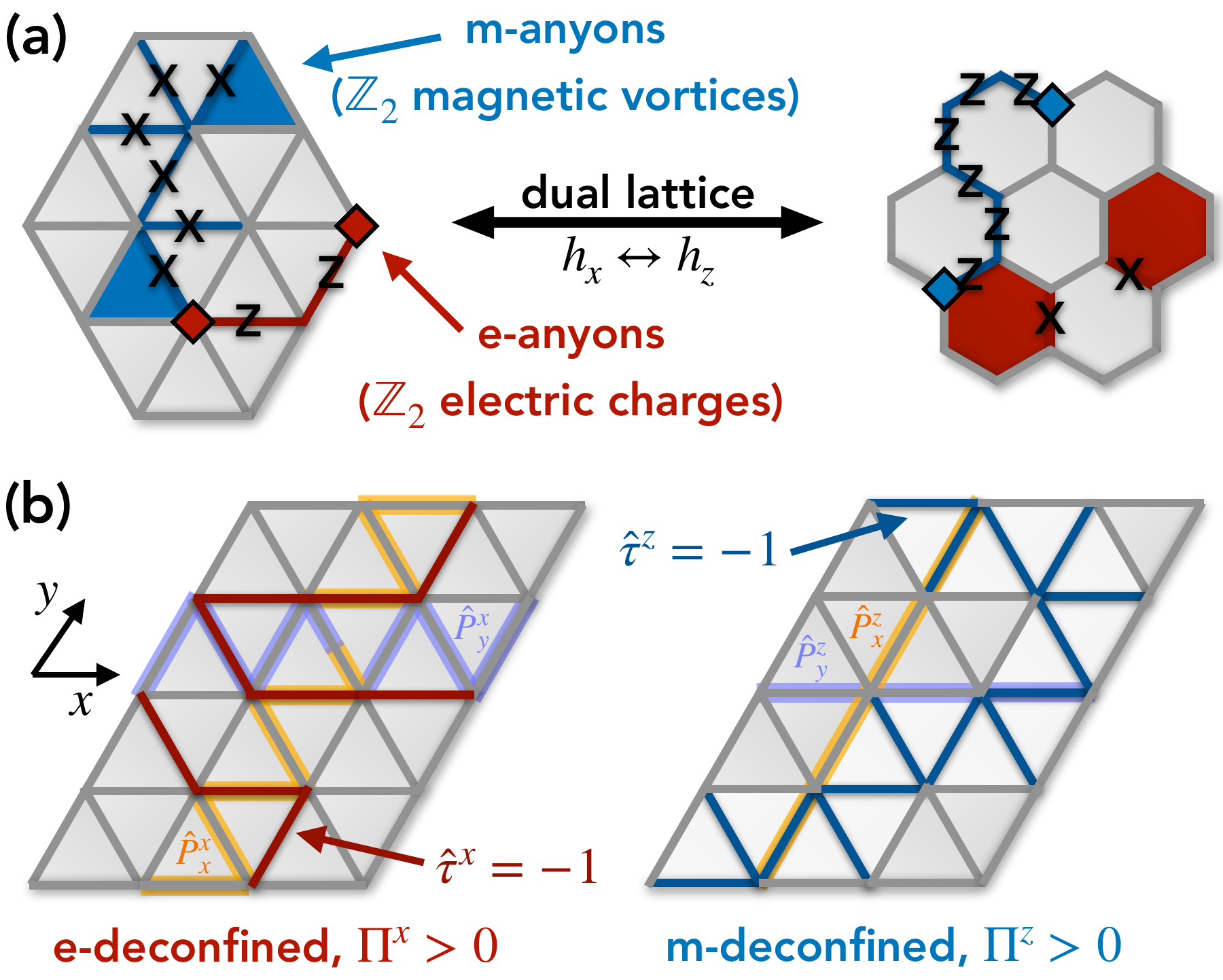}
\caption{\textbf{Link- and plaquette percolation} as confinement order parameters for e- and m-anyons in the extended toric code~(\ref{eq:eTC}). (a) On the left, we start with a triangular lattice in the ground state of the bare toric code ($h_x=h_z=0$), i.e. $\hat{A}_v=\hat{B}_p=1$ everywhere. Applying $\hat{\tau}^z$-perturbations (``Z'') on a cluster of neighboring links creates a pair of frustrated stars ($\hat{A}_v=-1$) at the open ends of the cluster, which are associated with point-like e-anyons living on lattice sites (red). Applying $\hat{\tau}^x$-perturbations (``X'') on a cluster of links connecting neighboring plaquettes creates a pair of frustrated plaquettes ($\hat{B}_p=-1$) at the open ends of the cluster, which are associated with point-like m-anyons living on plaquettes (blue). The problem can be mapped to the dual lattice (here: honeycomb lattice) where the meaning of e- and m-anyons is swapped along with the fields ($h_x \leftrightarrow h_z$). (b) In the e-deconfined phase a percolating cluster of links with $\hat{\tau}^x = -1$ winds around the periodic lattice in at least one spatial dimension (here $\hat{P}^x_x=\hat{P}^x_y=-1$), corresponding to a non-zero percolation probability $\Pi^x > 0$. In the m-deconfined phase a percolating plaquette cluster connected by links with $\hat{\tau}^z = -1$ winds around the periodic lattice in at least one spatial dimension (here $\hat{P}^z_x=1$, $\hat{P}^z_y=-1$), corresponding to a non-zero plaquette percolation probability $\Pi^z > 0$.}
\label{figAnyons}
\end{figure}

The bare toric code, i.e. $h_x=h_z=0$, features an exactly solvable topological ground state with $\hat{A}_v=\hat{B}_p=1 \; \forall v,p$ which is an equal superposition of closed loops of connected links with $\hat{\tau}^x=-1$ ($\hat{\tau}^z=-1$) on the (dual) lattice, typically referred to as a ``quantum loop gas''. The four topological sectors of the ground state can be equally distinguished in the $\hat{\tau}^x$- or $\hat{\tau}^z$-basis by the expectation values of non-contractible loop operators, measuring the winding number parity in $x$ or $y$-direction:
\begin{align}
    \hat{P}^x_{x/y} &= \prod_{l \in \tilde{\gamma}_{y/x}} \hat{\tau}^x_l, \\
    \hat{P}^z_{x/y} &= \prod_{l \in \gamma_{y/x}} \hat{\tau}^z_l, 
\end{align}
where $\gamma_{\alpha}$ ($\tilde{\gamma}_{\alpha}$) is a non-contractible loop winding around the $\alpha$-direction ($\alpha \in \{x, y\}$) of the (dual) lattice. We illustrate the winding number parity in Fig.~\ref{figAnyons}b. 

Two-dimensional toric codes feature two types of anyonic excitations (illustrated in Fig.~\ref{figAnyons}a): e-anyons ($\mathbb{Z}_2$ electric charges) are point-like excitations on a frustrated star ($\hat{A}_v=-1$). They can be created in pairs by applying $\hat{\tau}^z$-perturbations on neighboring links on the bare toric code ground state with $\hat{A}_v=\hat{B}_p=1$ everywhere, thus creating a cluster of links with $\hat{\tau}^x=-1$. m-anyons ($\mathbb{Z}_2$ magnetic vortices) are point-like excitations associated with a frustrated plaquette ($\hat{B}_p=-1$). They can be created in pairs by applying $\hat{\tau}^x$-perturbations on links connecting elementary plaquettes on the ground state, thus creating a plaquette cluster connected by links with $\hat{\tau}^z=-1$. For the bare toric code $h_x, h_z = 0$, anyons are thermal excitations of the ground state.

When performing a duality mapping \cite{Wegner1971}, i.e. identifying plaquettes with sites, e-anyons on a two-dimensional lattice can be identified with m-anyons on the dual lattice and vice versa, see Fig.~\ref{figAnyons}a. In the case of the self-dual Fradkin-Shenker model \cite{Fradkin1979}, i.e. the extended toric code on the square lattice, e-anyons behave identically to m-anyons upon changing $h_x \leftrightarrow h_z$. This is reflected in the naming of the model's two phases, the confined (topological) phase where both e- and m-anyons are confined (deconfined). Crucially, this is not the case for the toric code on the triangular and the honeycomb lattice which are instead dual to each other. Later we will show that e- and m-anyons are independently confined on the triangular and honeycomb lattices. 

Different classes of order parameters are commonly used in the literature to probe topological order. The first class are string-loop order operators derived from Wegner-Wilson \cite{Wegner1971, Wilson1974} and 't~Hooft loops \cite{Thooft1978}. A prominent example is the Fredenhagen-Marcu (FM) order parameter, which was introduced in the context of $\mathbb{Z}_2$ lattice gauge theories \cite{Fredenhagen1983, Fredenhagen1986, Fredenhagen1988}. Its equal-time variant is defined as 
\begin{align}
    O_\mathrm{FM}^{x/z} = \lim_{L \to \infty} \frac{\langle \prod_{l \in \mathcal{C}^{x/z}_{1/2}} \hat{\tau}^{x/z}_l \rangle}{\sqrt{|\langle \prod_{l \in \mathcal{C}^{x/z}} \hat{\tau}^{x/z}_l \rangle|}},
\end{align}
where $\mathcal{C}^{x/z}$ is a closed contour of links with perimeter $\mathcal{O}(L)$ at equal imaginary time on the lattice ($\hat{\tau}^z$-basis; ``Wegner-Wilson loop'') or the dual lattice ($\hat{\tau}^x$-basis; ``'t~Hooft loop''); $\mathcal{C}^{x/z}_{1/2}$ is an open contour with two open ends that contains half the links of $\mathcal{C}^{x/z}$. The FM order parameter measures the response of the system when two e-anyons (m-anyons) are spatially separated and relates it to the response of the bulk, thus circumventing the problem that for non-zero $h_z$ ($h_x$) regular Wegner-Wilson ('t~Hooft) loops follow a perimeter (area) law in both the trivial and the topological phase \cite{Sachdev2018, Xu2024}. Another variant that circumvents the above problem is locally error-corrected decoration (LED) \cite{Cong2024}, where a renormalization group procedure is used to remove anyons from snapshots before measuring the string-loop operators.

In the context of continuous-time QMC, a staggered imaginary times (SIT) order parameter local in space but non-local in imaginary time has shown some success \cite{Wu2012}. It can be defined as 
\begin{align}
    O_{\mathrm{SIT}}^{x/z} = &\frac{1}{\beta} \bigl[ ( \tau_1^k - 0 ) - ( \tau_2^k - \tau_1^k ) + ... + (-1)^{N(k)-1} \nonumber \\
    &( \tau_{N(k)}^k - \tau_{N(k)-1}^k ) + (-1)^{N(k)} ( \beta - \tau_{N(k)}^k ) \bigr],
\end{align}
where $\tau_n^k$ is the imaginary time of the $n$-th tuple spin flip of type $k$. $k$ is an elementary plaquette $p$ (star $s$) of links when sampling in the $\hat{\tau}^x$-basis ($\hat{\tau}^z$-basis)\footnote{It is vital that in the QMC the SIT is \textit{not} averaged over the whole lattice before evaluating the Binder cumulant, otherwise it shows crossover behavior for physical phase transitions and the Binder cumulant does not feature crossing points!}. For details, see \cite{Wu2012}. 

Recently, POPs have been proposed in the context of e-confinement in $\mathbb{Z}_2$ lattice gauge theories \cite{Linsel2024}. They measure the winding number of connected clusters $C$ of adjacent links $l$ with $\hat{\tau}^x_l = -1 \; \forall l \in C$. The physical intuition comes from the picture of fluctuating $\hat{\tau}^x$-fields which connect pairs of e-anyons in local clusters in the confined phase and form global percolating clusters in the deconfined (topological) phase, see Fig.~\ref{figAnyons}b. In its simplest form, the \textit{percolation probability} can be written as the expectation value $\langle \hat{\Pi}^x \rangle$ of the projector
\begin{align}
    \hat{\Pi}^x = \sum_{ W(j) \neq 0} \ket{\{\hat{\tau}^x\}_j} \bra{\{\hat{\tau}^x\}_j}
\end{align}
over all possible configurations $\{\hat{\tau}^x\}_j$ with non-zero winding number $W(j)$, i.e. configurations with zero winding number have $\Pi^x=0$. Pictorially, $\hat{\Pi}^x$ measures whether it is possible to traverse the system only on links $l$ with $\hat{\tau}^x_l = -1$, hence the name ``percolation probability''. Similar order parameters have been used in the context of quantum chromodynamics \cite{Patel1984, Satz1998, Ghanbarpour2022}.

%Similarly, the \textit{percolation strength} can be defined as the expectation value $\langle \hat{P}^x \rangle$ of 
%\begin{align}
%    \hat{P}^x = \hat{\Pi}^x \max_c |c| / N_{\mathrm{links}},
%\end{align}
%i.e. as the number of links $|C_\textrm{max}|$ in the largest cluster $C_\textrm{max}$ with $\hat{\tau}^x_l = -1 \; \forall l \in C_\textrm{max}$ normalized by the total number of links $N_{\mathrm{links}}$ given that the system percolates \cite{Homeier2022}.

Here, we not only measure percolation in the $\hat{\tau}^x$-basis to probe the confinement of e-anyons but also extend the definition, by analogy, to the $\hat{\tau}^z$-basis to measure the confinement of m-anyons. Instead of bond percolation, we calculate the \textit{plaquette percolation probability} $\Pi^z$ where two neighboring plaquettes $p_1$ and $p_2$ are part of the same cluster iff they share a link $l \in p_1,p_2$ with $\hat{\tau}^z_l = -1$, see Fig.~\ref{figAnyons}b. 

For the quantum loop gas, i.e. the four-fold degenerate ground state of the bare toric code, only non-contractible, percolating loops lead to $\hat{P}^x_{x/y} = -1$ ($\hat{P}^z_{x/y} = -1$) since all contractible loops contribute an overall factor of $+1$ to the winding number parity by construction, see Fig.~\ref{figAnyons}b. The topological sector with $\hat{P}^x_{x} = \hat{P}^x_{y} = 1$ ($\hat{P}^z_{x} = \hat{P}^z_{y} = 1$) is compatible with percolating loops, hosting e.g. two percolating clusters or one percolating cluster with an even, non-zero winding number. Thus the topological quantum loop gas has the necessary condition $\Pi^x, \Pi^z > 0$ as a \textit{direct consequence of its topological degeneracy}, underlining that percolation is a very natural quantity to probe in the context of topological phases. We will demonstrate that this conjecture also holds for the toric code subject to a finite external field. Even the absence of percolation in one basis, i.e. the confinement of either e- or m-anyons, already implies a topologically trivial phase.

We will measure and compare the FM, SIT and POPs to gain insights into the ground state phase diagram of the extended toric code~(\ref{eq:eTC}) on various lattices. All three order parameters are basis-dependent and it is practically impossible to reproduce results from the $\hat{\tau}^x$-basis in the $\hat{\tau}^z$-basis and vice versa, reflecting the absence of a local order parameter. Therefore, we will study both and combine their total information to understand the phase diagrams.

\begin{figure*}[t]
\includegraphics[width=0.95\textwidth]{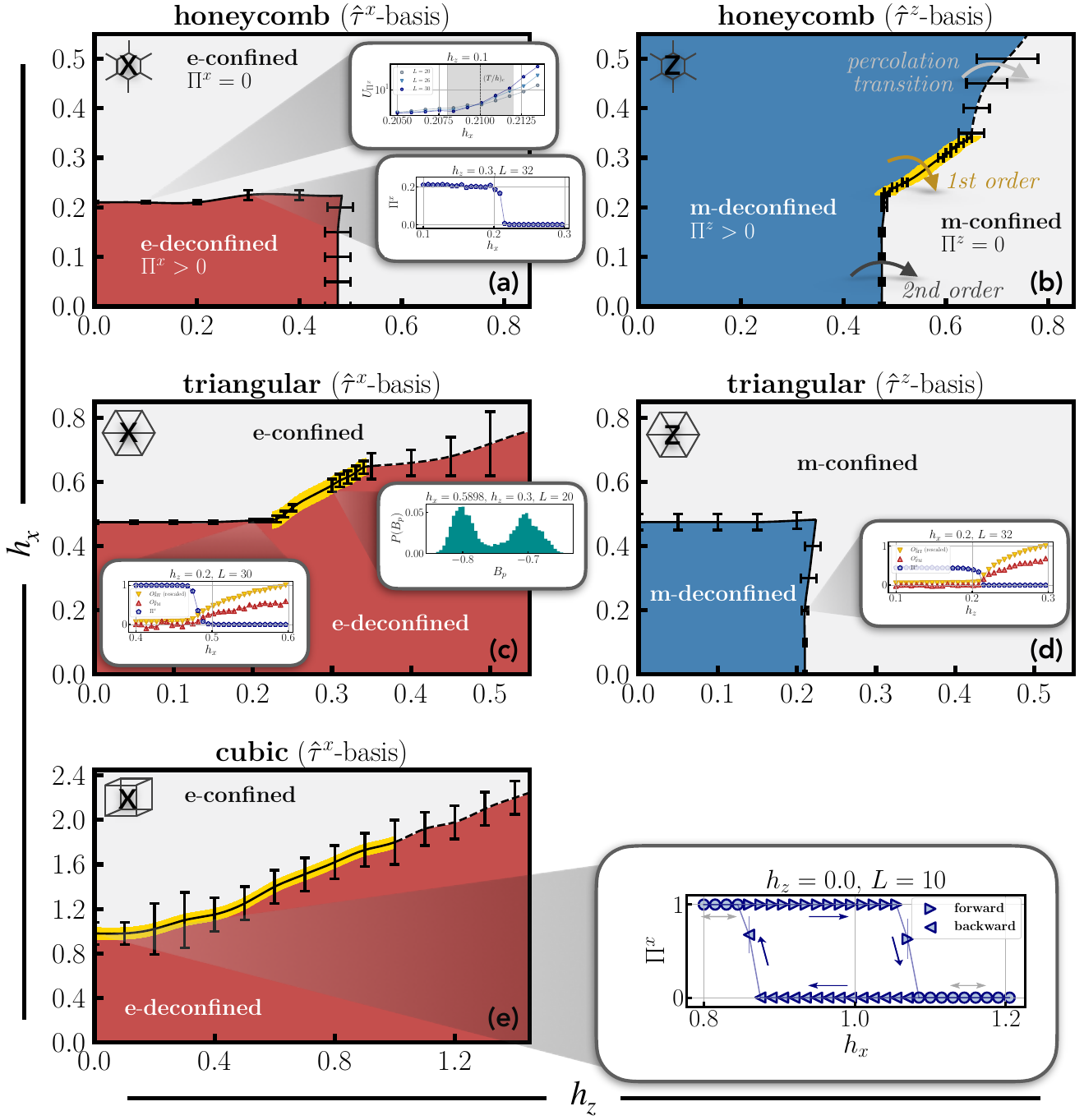}
\caption{\textbf{e- and m-confinement phase diagram} of the extended toric code~(\ref{eq:eTC}) on the honeycomb, triangular and cubic lattices. We apply continuous-time QMC at temperature $T=1/L$ to gain insights into the ground state phase diagram and use a crossing-point analysis of Binder cumulants $U$ to extract the critical fields, see inset of (a). On the left (right), we sample in the $\hat{\tau}^x$-basis ($\hat{\tau}^z$-basis) and measure the percolation probability $\Pi^x$ (plaquette percolation probability $\Pi^z$), the FM and the SIT order parameters. (a-b) The honeycomb lattice features an extended e-deconfined region with $\Pi^x>0$ for small $h_x, h_z$ and an e-confined region with $\Pi^x=0$ for larger $h_x, h_z$ which are separated by a continuous phase transition. m-anyons are always deconfined ($\Pi^z>0$) for small $h_z$, giving rise to an e-confined, m-deconfined regime. For large $h_z$, m-anyons feature a confined phase. The phase boundary features two multi-critical points between which the phase transition is of first-order type. For even higher fields, $\Pi^z$ has a percolation transition (dashed line) which is not relevant for topology. (c-d) The toric code on the triangular lattice is dual to the honeycomb lattice, the phase diagram is identical to the honeycomb lattice when exchanging the basis $\hat{\tau}^x \leftrightarrow \hat{\tau}^z$ and $h_x \leftrightarrow h_z$. We identify an m-confined, e-deconfined regime for large $h_z$. (e) On the cubic lattice, the confinement of e-anyons is qualitatively similar to the triangular lattice but features a first-order phase transition (yellow phase boundary, see inset for hysteresis curve) between the e-deconfined (e-confined) region at small (large) $h_x$. The first order line ends at a multi-critical point around $(h_z, h_x) = (1.0(1), 1.8(2))$ after which we find a percolation transition (dashed line). (a-e) We identify the topological phase with the regime where both e- and m-anyons are deconfined. The FM and SIT order parameters can generally probe the topological phase transition in the $\hat{\tau}^z$-basis ($\hat{\tau}^x$-basis) for $h_z$-scans ($h_x$-scans), see insets in c, d. In the other two cases, i.e. in the $\hat{\tau}^x$-basis ($\hat{\tau}^z$-basis) for $h_z$-scans ($h_x$-scans) the SIT features crossover behavior and the FM is too noisy to be evaluated. Additional data is presented in Appendix~\ref{app:details}.}
\label{figConfinementPD}
\end{figure*} 

\section{Phase diagrams}

\subsection{Honeycomb lattice}

On the honeycomb lattice, we simulate periodic systems up to $L^2=32^2$ in terms of unit cells at $T=1/L$ and take up to $3 \times 10^4$ snapshots for every data point. A recent work \cite{Kott2024} provides a good starting point for interesting parameter ranges. We repeat our parameter scans in both the $\hat{\tau}^x$- and $\hat{\tau}^z$-bases and calculate the FM, SIT, POPs and other observables like the energy for every QMC snapshot. Here and in the following, all error bars are calculated using either the integrated autocorrelation time or the stationary bootstrap \cite{Politis1994, Politis2004, Patton2009}. 

We show the e-confinement phase diagram in Fig.~\ref{figConfinementPD}a. We observe an extended e-deconfined phase with $\Pi^x>0$ that is persistent for finite fields $h_x, h_z > 0$ and an e-confined phase with $\Pi^x=0$ for large fields. For $h_x$-scans at small $h_z$, we find a continuous phase transition to the e-confined regime. At $h_z=0$, the model can be exactly mapped to the transverse-field Ising model (TFIM) on the \textit{triangular} lattice \cite{Kott2024}, whose phase transition is known to be in the (2+1)D Ising universality class (implying (2+1)D Ising* for the extended toric code). Our critical field $h_{x,\mathrm{c}} = 0.210(2)$ is in good agreement with TFIM QMC studies which yield $h^{\mathrm{TFIM}}_{x,\mathrm{c}} \approx 0.209$ \cite{Blote2002}. We repeat the finite-size scaling for the SIT and the results agree with the POPs. The FM order parameter does capture the phase transition from the e-deconfined phase into the e-confined phase ($h_x \gg 1$) but not into the Higgs phase ($h_z \gg 1$), where it is too noisy to be evaluated. The QMC sampling in the $\hat{\tau}^x$-basis becomes vastly inefficient for high $h_z$ because spins are more aligned with the perpendicular $\hat{\tau}^z$-operator. As a result, the observables in this range tend to be noisy and a high number of snapshots is required. For larger fields $h_z \geq 0.3$, the POP Binder cumulants do not show a crossing point for the simulated system sizes, but we still observe a clear drop to $\Pi^x=0$ for large fields $h_x, h_z$, see inset. Similarly, for $h_z$-scans we do not find SIT Binder cumulant crossing points for the simulated system sizes, but we observe an abrupt rise in $O^\mathrm{SIT}$ at the critical fields (known from the duality to the triangular lattice). We also study a classical limit of the extended toric code (\ref{eq:eTC}) on the honeycomb lattice in Appendix \ref{app:classical_mc_sampling}. 

In Fig.~\ref{figConfinementPD}b, we show the m-confinement phase diagram. For $h_z$-scans, we observe a transition from an m-deconfined regime with $\Pi^z > 0$ to an m-confined regime with $\Pi^z = 0$. At $h_x=0$ the critical value is $h_{z,\mathrm{c}} = 0.475(5)$ obtained by a crossing-point analysis. This value is confirmed by the SIT and FM order parameters. In stark contrast to the e-confinement, we do not observe a transition into an m-confined regime for large $h_x$ but small $h_z$. The system remains m-deconfined and $\Pi^z>0$, while it is e-confined and $\Pi^x=0$, hence, e- and m-anyons can be independently confined at zero temperature. For details, see Appendix~\ref{app:details}.

Another interesting feature of the m-confinement phase diagram in Fig.~\ref{figConfinementPD}b is the existence of a first-order phase boundary between the m-deconfined and m-confined phase which starts at a multi-critical point around the tip of the e-deconfined phase in Fig.~\ref{figConfinementPD}a at $(h_z, h_x) = (0.485(5),0.225(5))$ and ends at a multi-critical point $(h_z, h_x) = (0.61(1),0.34(1))$ after which we find a percolation transition (dashed line). This transition is a feature of POPs and important for confinement, but crucially does not signal a topological phase transition, i.e. it is compatible with the Fradkin-Shenker theorem \cite{Fradkin1979}. The first-order nature is signaled by a double-peak structure of the histogram of the probability distribution of observables obtained from QMC snapshots, indicating the coexistence of two phases. We find the double-peak structure not only in $\Pi^z$ but also in the energy which is a clear sign that this first-order line is indeed physical. An exemplary histogram is shown in the inset of Fig.~\ref{figConfinementPD}c. Crucially, the first-order line is not visible in $\Pi^x$.

%For infinite fields, we expect a Dicke transition \cite{Dicke1954, Hepp1973} between the fully polarized states with $\hat{\tau}^{x/z}_l = 1 \, \forall l$, respectively.

\begin{figure}[]
\includegraphics[width=0.45\textwidth]{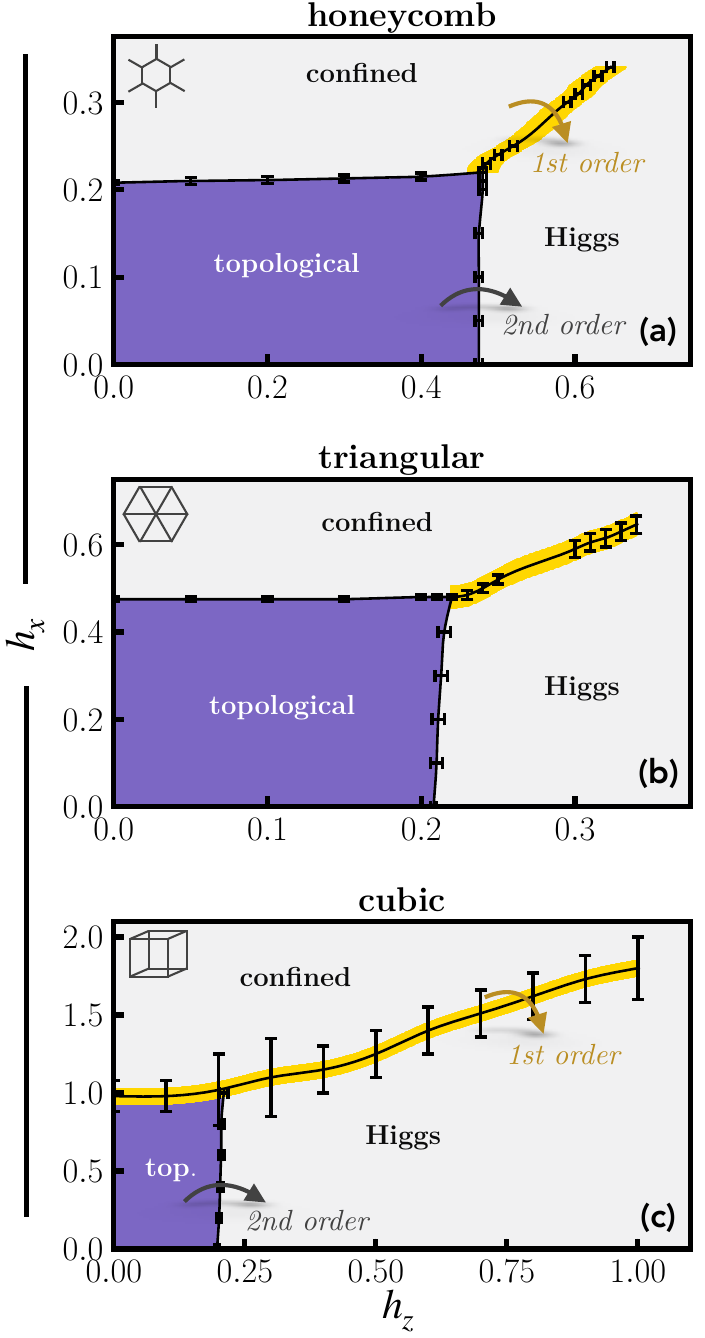}
\caption{\textbf{Topological phase diagram} of the extended toric code~(\ref{eq:eTC}) on the honeycomb, triangular and cubic lattices. We construct the topological phase diagram using ground state QMC calculations for the FM, SIT and POPs which are partially shown in Fig.~\ref{figConfinementPD}. The general structure of the phase diagram is remarkably similar to the Fradkin-Shenker model \cite{Fradkin1979} for all lattices: the system features an extended topological phase that is persistent for finite fields $h_x, h_z > 0$. The critical line $h_{x, \mathrm{c}}(h_z)$ ($h_{z,\mathrm{c}}(h_x)$) - extracted using a finite-size scaling analysis - is convex, i.e. it slightly shifts up for increasing $h_z$ ($h_x$). A first-order line starts at a multi-critical point at the tip of the topological phase and ends at a multi-critical point in the trivial phase. (a) honeycomb lattice: a continuous phase transition separates the topological from the trivial regime ($h_x, h_z \gg 1$). (b) triangular lattice: the honeycomb lattice is dual to the triangular lattice, which is reflected in the topological phase diagram ($h_x \leftrightarrow h_z$). (c) cubic lattice: the system features a first-order phase transition as we increase $h_x$ while fixing $h_z$ (yellow phase boundary). The phase boundary where $h_x$ is kept constant is associated with a continuous phase transition (solid line).}
\label{figTopology}
\end{figure}

We identify the topological phase with the phase where both e- and m-anyons are deconfined and show the resulting topological phase diagram on the honeycomb lattice in Fig.~\ref{figTopology}a. The phase diagram is constructed using the FM, SIT and POPs in the \mbox{$\hat{\tau}^x$-} \mbox{($\hat{\tau}^z$-)} basis for \mbox{$h_x$-} \mbox{($h_z$-)} sweeps and all order parameters agree. It is reminiscent of previous studies \cite{Kott2024} and qualitatively similar to the square lattice \cite{Wu2012}. Interestingly, it exhibits a first-order line akin to the Fradkin-Shenker phase diagram.

\subsection{Triangular lattice}

Our approach on the triangular lattice is similar to the honeycomb lattice. We simulate periodic systems up to $L^2=32^2$ in terms of unit cells at $T=1/L$ and take up to $3 \times 10^4$ snapshots for every data point. We use the duality between the triangular and honeycomb lattice toric code to identify interesting parameter regimes. We repeat our parameter scans in both the $\hat{\tau}^x$- and $\hat{\tau}^z$-bases and calculate the FM, SIT, POPs and other observables like the energy for every QMC snapshot. 

We show the e-confinement phase diagram in Fig.~\ref{figConfinementPD}c. Due to the duality to the honeycomb lattice, it is identical to Fig.~\ref{figConfinementPD}b when exchanging $h_x \leftrightarrow h_z$. We observe an extended e-deconfined phase with $\Pi^x>0$ that is persistent for finite fields $h_x, h_z > 0$. For $h_x$-scans, we observe a continuous phase transition to an e-confined regime where $\Pi^x = 0$. The critical fields have been extracted using a Binder cumulant crossing-point analysis. At $h_z=0$, the model can be exactly mapped to the TFIM on the \textit{honeycomb} lattice \cite{Wegner1971}, whose phase transition is in the (2+1)D Ising universality class. Our critical field $h_{x,\mathrm{c}} = 0.475(5)$ is in good agreement with TFIM QMC studies which yield $h^{\mathrm{TFIM}}_{x,\mathrm{c}} \approx 0.469$ \cite{Blote2002}. We repeat the finite-size scaling for the SIT and the results agree with the POPs (see inset for FM and SIT results). As on the honeycomb lattice, we observe the first-order line between the multi-critical points $(h_z, h_x) = (0.225(5), 0.485(5))$ and $(h_z, h_x) = (0.34(1), 0.61(1))$. 

For even higher $h_z$, we find a percolation transition (dashed line). We can explain this transition by looking at the limit $h_x=0, h_z \to \infty$ where all spins are magnetized in $\hat{\tau}^z$-direction, see Appendix~\ref{app:details}. In the $\hat{\tau}^x$-basis, the spins are completely random and independent, i.e. we have Bernoulli bond percolation with a probability $p=0.5$ which is larger than the Bernoulli bond percolation threshold on the triangular lattice, $p_{\mathrm{c,tri}} \approx 0.35$ \cite{Sykes1964, Wierman1981}, hence $\Pi^x > 0$. In the other limit $h_z=0, h_x \to \infty$, all spins are magnetized in $\hat{\tau}^x$-direction, hence $\Pi^x = 0$. The dashed line is the percolation transition between these two limits. On the cubic lattice, the Bernoulli bond percolation threshold is $p_{\mathrm{c,cu}}=0.247(5)$ \cite{Sykes1964}, indicating a similar situation as on the triangular lattice, see discussion Fig.~\ref{figConfinementPD}e. On the honeycomb lattice, however, the Bernoulli bond percolation threshold is $p_{\mathrm{c,hon}} \approx 0.65$ \cite{Sykes1964, Wierman1981}, thus both limits are e-confined, confirming our numerical result in Fig.~\ref{figConfinementPD}a.  

In Fig.~\ref{figConfinementPD}d, we show the m-confinement phase diagram. The phase diagram is identical to the e-confinement phase diagram of the honeycomb lattice when exchanging $h_x \leftrightarrow h_z$. An extended m-deconfined phase for small fields $h_x, h_z > 0$ has a continuous phase transition to an m-confined phase for larger fields. For large $h_z$, the system is e-deconfined but m-confined. The SIT Binder cumulant crossing points confirm the phase boundaries of $\Pi^z$ (see inset for FM and SIT results). The (physical) first-order line cannot be probed using $\Pi^z$.

We show the resulting topological phase diagram on the triangular lattice in Fig.~\ref{figTopology}b. The duality between the toric code on the triangular and honeycomb lattice is reflected in the topological phase diagram ($h_x \leftrightarrow h_z$).

%Fradkin and Shenker \cite{Fradkin1979} famously proved for the square lattice that the confined ($h_x \gg 1$) and Higgs phase ($h_z \gg 1$) are analytically connected and thus identical. We expect a similar theorem to hold here. However, $\Pi^x$ and $\Pi^z$ show critical behavior outside the topological phase. We believe this is an unphysical manifestation of Kolmogorov's zero-one-law \cite{Lyons1990}, which states that the percolation probability $\Pi$ can only be exactly zero or exactly one in the thermodynamic limit, thus ruling out crossover behavior. In the future, it would be interesting to adapt the POPs in a way such that they can in principle host crossover behavior, e.g. by including the size of the \textit{second-largest} cluster.

\subsection{Cubic lattice}

We study the cubic lattice toric code with plaquette interactions, i.e. the interaction of 4 links on cube faces but crucially \textit{not} with cubic interactions of 12 links. The star interaction is a 6-qubit term involving all links in three spatial dimensions connected to a given site. In general, the correct definition of a POP in three dimensions heavily depends on the details of the model. E.g., cube interactions lead to a different excitation structure than plaquette interactions. For this reason, we do not study m-confinement with percolation but instead rely on the SIT and FM to map out the topological phase boundaries. We simulate periodic systems up to $L^3=16^3$ in terms of unit cells at $T=1/L$ and take up to $10^4$ snapshots for every data point. An earlier work \cite{Reiss2019} provides a good starting point for interesting parameter ranges. We repeat our parameter scans in both the $\hat{\tau}^x$- and $\hat{\tau}^z$-bases and calculate the FM, SIT, POPs and other observables like the energy for every QMC snapshot. 

We show the e-confinement phase diagram in Fig.~\ref{figConfinementPD}e. We detect an extended e-deconfined phase with $\Pi^x>0$ that is persistent for finite fields $h_x, h_z > 0$. For $h_x$-scans, we observe a first-order phase transition to an e-confined regime, where $\Pi^x = 0$. To obtain the hysteresis curve, we used the state from the previous data point as the initial state, respectively. We observe an enormous hysteresis area (yellow phase boundaries), which we show for $h_z=0$ in the inset. At $h_z=0$, the model is equivalent to Wegner's 4D lattice gauge theory \cite{Wegner1971} which features a first-order transition \cite{Balian1975, Creutz1979} around the critical field $h^{\mathrm{self-duality}}_{x,\mathrm{c}} = 1$ \cite{Kott2024} obtained from self-duality. Our critical field $h_{x,\mathrm{c}} = 0.98(10)$ (the error bars are the hysteresis interval) is consistent with this result. The hysteresis is also visible in other observables like energy. The first-order phase boundary ends at a multicritical point $(h_z, h_x) = (1.0(1), 1.8(2))$. For even higher fields, $\Pi^x$ has a percolation transition (dashed line).

For $h_z$-scans at small $h_x$, the system remains e-deconfined. However, it features a phase transition to the topologically trivial phase signaled by the SIT and FM in the $\hat{\tau}^z$-basis. At $h_x=0$, the model can be exactly mapped to the TFIM on the \textit{cubic} lattice \cite{Zarei2017}, whose phase transition is in the (3+1)D Ising universality class (implying (3+1)D Ising* for the extended toric code). The critical field $h^{\mathrm{TFIM}}_{z,\mathrm{c}} \approx 0.194$ \cite{Jansen1989} is in good agreement with our SIT result $h_{z,\mathrm{c}} = 0.197(05)$.

We show the resulting topological phase diagram for the cubic lattice in Fig.~\ref{figTopology}c. It is structurally identical to the Fradkin-Shenker phase diagram except for one crucial difference: It features a first-order phase transition between the topological and trivial phase for small $h_z$ which leads to strong hysteresis in all observables. The topological phase diagram also features a multi-critical point at $(h_z, h_x) = (0.210(08), 1.0(2))$, the tip of the topological phase, and at $(h_z, h_x) = (1.0(1), 1.8(2))$, the end of the first-order line.

\subsection{Order parameter comparison}

Using the SIT, we can reliably perform a finite-size scaling analysis for the topological phase transition for $h_z$-scans ($h_x$-scans) in the $\hat{\tau}^z$-basis ($\hat{\tau}^x$-basis). In the other two cases, i.e. $h_x$-scans ($h_z$-scans) in the $\hat{\tau}^z$-basis ($\hat{\tau}^x$-basis), the SIT shows crossover behavior but it is crucially able to signal the rough phase boundary. However, it is inaccessible not only to other numerical methods without easy access to imaginary time, like tensor networks, but also to experiments. In its current definition in imaginary time, it further lacks a clear physical meaning. Like for the other order parameters, it is necessary to measure both in the $\hat{\tau}^x$- and the $\hat{\tau}^z$-basis to perform a crossing-point analysis for all topological phase boundaries, requiring separate QMC snapshots, respectively.

In contrast to the SIT, both the FM and the POPs are accessible to quantum simulators. Both can reliably probe the topological phase transition for $h_z$-scans ($h_x$-scans) in the $\hat{\tau}^z$-basis ($\hat{\tau}^x$-basis). However, the parameter range for the finite-size scaling is typically larger for the POPs, as the FM is a ratio of two exponentially small numbers which generally results in large statistical errors even for small fields $h_x, h_z$. In addition, the FM exhibits extreme levels of noise for $h_x$-scans ($h_z$-scans) in the $\hat{\tau}^z$-basis ($\hat{\tau}^x$-basis), rendering it practically unusable. In a recent tensor network study \cite{Xu2024}, the FM was further found to host unphysical singularities in the flux-condensing confined region ($h_x \gg 1$). The exact form of the string-loop operator has to be adapted to the problem at hand. Similarly, the POP has to be adapted for other Hamiltonians and lattices, too (e.g. for the toric code on the three-dimensional cubic lattice). 

All order parameters studied in this work are basis-dependent and no single order parameter captures all phase boundaries in its Binder cumulant using only one basis. Vice versa, given one basis and all order parameters in that basis, it is impossible to faithfully calculate the full phase diagram using Binder cumulants thus far, underlining the importance of measuring in different bases. The SIT is more robust than the FM and POPs, however, the SIT is not accessible to quantum simulators, ruling it out for experimental studies.  

%%%%%%%%%%%%%%%%%%%%%%%%
% Discussion and outlook
%%%%%%%%%%%%%%%%%%%%%%%%

\section{Discussion and outlook}

We mapped out the phase diagrams of the extended toric code on the triangular, honeycomb and cubic lattices using numerically exact continuous-time quantum Monte Carlo simulations, precisely determining their phase boundaries and the order of their transitions, going beyond previous work~\cite{Reiss2019, Kott2024}. Our work manifestly demonstrates that, even in the ground state, we must make a distinction between topological order and \mbox{(de-)confinement}. Whereas topological order coincides with deconfinement of e- and m-anyons together, the probing of confinement of e- and m- anyons separately depends on the choice of basis and quantity of interest, a situation which is exacerbated for lattices that are not self-dual. In particular, choosing percolation, which has the advantage of being experimentally accessible in current-generation quantum simulators~\cite{Zohar2017, Barbiero2019, Schweizer2019_2, Homeier2021, Homeier2022, Mildenberger2025, Halimeh2025}, we see that e-confinement agrees with a topologically trivial state on the honeycomb lattice in the $\hat{\tau}^x$-basis. The same is true for m-confinement on the triangular lattice in the $\hat{\tau}^z$-basis. In contrast, m-deconfinement on the honeycomb lattice in the $\hat{\tau}^z$-basis is found over a much larger area in the phase diagram compared to the topological order (and the same is true for the triangular lattice in the $\hat{\tau}^x$-basis). Other quantities, such as the SIT, which is susceptible to dynamical effects, lead to qualitatively similar but quantitatively different behavior. Conceptually similar issues were previously reported for the FM order parameter~\cite{Xu2024}. All these observations are ultimately related to the nonexistence of a local order parameter for such Ising-like transitions. Note that our results remain fully compatible with the Fradkin-Shenker argument. Our work also connects to the recently popular $PXP$ model on the ruby lattice with an emergent odd $\mathbb{Z}_2$ lattice gauge theory where e- and m-deconfinement were likewise found in topologically trivial phases \cite{Verresen2021, Wang2025}.

We leave for future work the detailed understanding of the universality class at the multi-critical points, which recently gathered a lot of attention on the square lattice, and a more detailed study of the SIT to probe the dynamical aspects of (de-)confinement.\medskip

%%%%%%%%%%%%%%%%%%%%%%%%%%%%%%%%%%%%%%%%%%%%%%%%%%%%%
\section*{Acknowledgments}

We thank G. Dünnweber and L. Homeier for fruitful discussions. We used the open-source library \textsc{ParaToric} for the continuous-time QMC simulations \cite{Linsel2025_2}. This research was funded by the European Research Council (ERC) under the European Union’s Horizon 2020 research and innovation program -- ERC Starting Grant SimUcQuam (Grant Agreement No. 948141) and QuantERA II (Grand Agreement No. 101017733), by the QuantERA grant DYNAMITE, and by the Deutsche Forschungsgemeinschaft (DFG, German Research Foundation) under project number 499183856 and under Germany's Excellence Strategy -- EXC-2111 -- project number 390814868.

\FloatBarrier

%%%%%%%%%%%%%%%%%%%%%%%%%%%%%%%%%%%%%%%%%%%%%%%%%%%%%
\section*{Appendix} \label{appendix}
\setcounter{section}{0}

%%%%%%%%%%%%%%%%%%%%%%%%%%%%%%%%%%%%%%%%%%%%%%%%%%%%%
\subsection{Classical limit of the toric code}
\label{app:classical_mc_sampling}
%%%%%%%%%%%%%%%%%%%%%%%%%%%%%%%%%%%%%%%%%%%%%%%%%%%%%

\begin{figure}[t]
\includegraphics[width=0.49\textwidth]{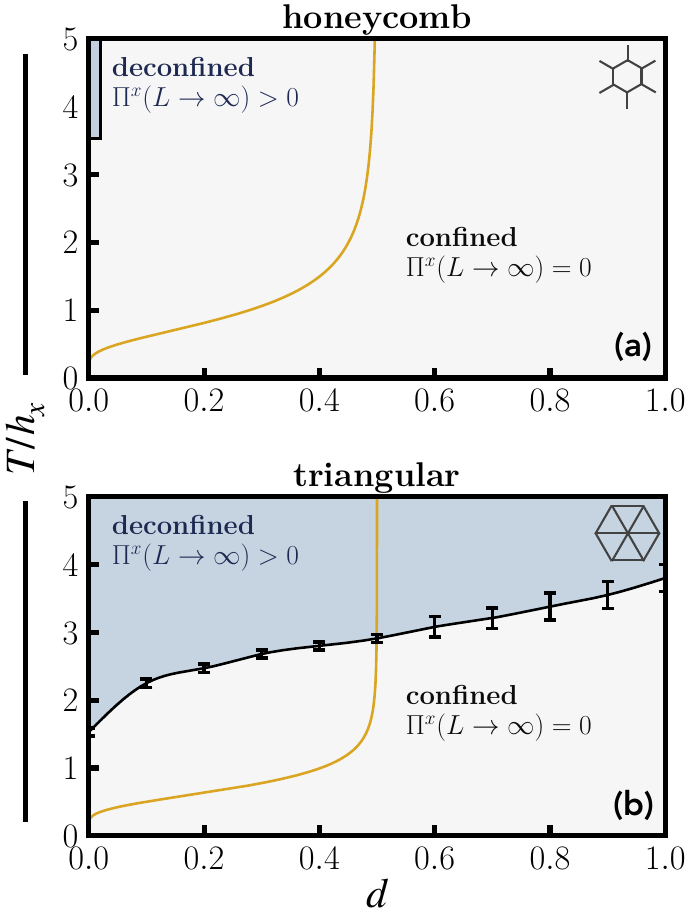}
\caption{\textbf{Classical confinement phase diagram} of Hamiltonian (\ref{eq:can_cl_ham}) at finite matter density $d$ on the honeycomb and triangular lattices. At zero matter density, the model can be mapped to the two-dimensional classical Ising model on the dual lattice \cite{Wegner1971}, respectively. The golden lines correspond to Hamiltonian (\ref{eq:gc_cl_ham}) at $\mu=0$. We use the percolation probability $\Pi^x$ in the $\hat{\tau}^x$-basis as a confinement order parameter. (a) honeycomb lattice: we find a thermal deconfinement transition at zero matter density. The critical temperature is the \textit{triangular} lattice Ising critical temperature $(T/h_x)_\mathrm{c} \approx 3.64$ \cite{Fisher1967}. At finite matter densities, the system is always non-percolating. (b) triangular lattice: the critical temperature at zero density is the \textit{honeycomb} lattice Ising critical temperature $(T/h_x)_\mathrm{c} \approx 1.52$ \cite{Fisher1967}. In contrast to the honeycomb lattice, we observe an extended region with $\Pi^x > 0$ for finite matter densities. The cubic lattice phase diagram is qualitatively similar to the triangular lattice  \cite{Linsel2024}.}
\label{fig:classical_pd}
\end{figure}

We perform classical Monte Carlo simulations of a classical limit of the extended toric code (\ref{eq:eTC})
\begin{align} \label{eq:can_cl_ham}
    \hat{\mathcal{H}}_{\mathrm{can}} = - h_x \,\hat{\mathcal{P}}\, \Bigl[\, \sum_{l} \hat{\tau}_{l}^x \,\Bigr] \,\hat{\mathcal{P}}\,
\end{align}
at a finite temperature $T/h_x$ and $h_x>0$. We define hard-core bosons $\hat{n}_{\j}$ on lattice sites $\j$. The projector $\hat{\mathcal{P}}$ fixes the total number of hard-core bosons $N=\sum_{\j} \hat{n}_{\j}$. This highly non-trivial model has deep connections to Hamiltonian (\ref{eq:eTC}) and has been studied numerically \cite{Linsel2024} and analytically \cite{Duennweber2025} in earlier studies. An interesting feature is the local $\mathbb{Z}_2$ gauge symmetry with the generator
\begin{align}
    \Gj = (-1)^{\nj} \prod_{l \in +_{\j}}\hat{\tau}_{l}^x. \label{eq:Goperator}
\end{align}
The property $[\hat{\mathcal{H}}_{\mathrm{can}}, \Gj] = 0$ directly results in a set of locally conserved eigenvalues $\Gj \ket{\Psi} = g_{\j} \ket{\Psi}$ (``background charges'') which we set to $g_{\j}=+1 \, \forall \j$, defining a so called ``gauge sector''. The eigenvalue equation is known as Gauss's law and restricts the physical Hilbert space. Note that the physics of the model depends on the choice of these background charges, it is not to be confused with a gauge transformation that leaves the physics unchanged.

In this canonical formulation, the density of matter particles $d$ is externally fixed. Matter can be introduced into the toric code by identifying an open end of a cluster $\Sigma$ of neighboring links with $\hat{\tau}^x_l = -1 \; \forall l \in \Sigma$, i.e. a site with an uneven number of $\hat{\tau}^x= -1$ attached to it, with a hard-core boson. 

Hamiltonian (\ref{eq:can_cl_ham}) can equally be formulated in a grand-canonical form where the density of bosons is controlled via a chemical potential $\mu$:
\begin{align} \label{eq:gc_cl_ham}
    \hat{\mathcal{H}}_{\mathrm{gc}} &= - h_x \sum_{l} \hat{\tau}_{l}^x - \mu \sum_{\j} \hat{n}_{\j} \nonumber \\
    &= - h_x \sum_{l} \hat{\tau}_{l}^x - \mu \sum_{\j} \frac{1}{2} \Bigl( 1- \prod_{l \in +_{\j}}\hat{\tau}_{l}^x \Bigr).
\end{align}
Note that the chemical potential term resembles the star term in the extended toric code (\ref{eq:eTC}). 

%Strictly speaking, the extended toric code is not a $\mathbb{Z}_2$ lattice gauge theory since the term $\propto \hat{\tau}^z$ explicitly violates the gauge symmetry, but can be mapped to one in the hard-core limit \cite{Linsel2024}.

At $\mu=0$, crucially without fixing the matter density, the probability $p$ for a given link $l$ to have $\hat{\tau}^x_l=-1$ is given by 
\begin{align}
    p = e^{-\beta h} / [2\cosh(\beta h)],
\end{align}
independently of other links, thus reducing to a Bernoulli percolation problem. An important observation is that $p \to 1/2$ as $\beta \to 0$, independent of the underlying lattice.

According to Gauss's law, matter particles are connected to a cluster $\Sigma$ of neighboring links with $\hat{\tau}^x_l = -1 \; \forall l \in \Sigma$, where the energy cost $2h_x\ell$ grows linearly with the number of neighboring links $\ell$ in the cluster (note that clusters can also form closed loops without any matter). At low temperatures, matter particles form mesonic pair states, where matter particles on neighboring sites are connected by a cluster of size one (``confined phase''). At higher temperatures, as a consequence of the competition between energy and entropy, a global cluster that winds around periodic boundaries can form, where matter particles are incoherent, free \Ztwo{}~charges (``deconfined phase'') \cite{Hahn2022}. This transition is associated with a percolation transition from a non-percolating confined regime at low temperatures to a deconfined regime at high temperatures \cite{Linsel2024}. As a finite-$T$ phase in two dimensions, the deconfined phase is not topological but connects to the $T=0$ topological phase of the toric code (which is also deconfined).

We simulate the periodic honeycomb lattice with system size up to $L^2=40^2$ and take up to $10^4$ snapshots. The phase diagram is shown in Fig.~\ref{fig:classical_pd}a. At zero matter density, the model can be mapped to the two-dimensional classical Ising model on the dual lattice \cite{Wegner1971}, i.e. the triangular lattice. The critical temperature is the \textit{triangular} lattice Ising critical temperature $(T/h_x)_\mathrm{c} \approx 3.64$ \cite{Fisher1967}. This critical temperature is confirmed by the Monte Carlo, where we observe thermal deconfinement and a percolating phase for high temperatures. At a non-zero matter density, we do not find a thermal deconfinement transition in the thermodynamic limit, i.e. the presence of matter prohibits the formation of a percolating cluster. This behavior can be easily understood for $\mu=0$ (golden line in Fig.~\ref{fig:classical_pd}a), where $p \to 1/2$ for $T/h_x \to \infty$, thus never reaching the Bernoulli (bond) percolation threshold on the honeycomb lattice, $p_{\mathrm{c,hon}} \approx 0.65$ \cite{Sykes1964, Wierman1981}. The phase diagram is structurally identical to the one of the square lattice \cite{Linsel2024}, where the structure of the phase diagram was confirmed using an analytical renormalization group study \cite{Duennweber2025}.

The phase diagram on the triangular lattice is shown in Fig.~\ref{fig:classical_pd}b. We simulate periodic systems up to $L^2=40^2$ and take up to $10^4$ snapshots. At zero matter density, the critical temperature is the \textit{honeycomb} lattice Ising critical temperature $(T/h_x)_\mathrm{c} \approx 1.52$ \cite{Fisher1967}. In contrast to the honeycomb lattice, we find a thermal deconfinement phase transition and thus a deconfined phase at non-zero matter density. We extract the critical temperatures using a finite-size scaling analysis. Looking again at the $\mu=0$ (golden line in Fig.~\ref{fig:classical_pd}b), we have $p \to 1/2$ for $T/h_x \to \infty$, thus reaching the Bernoulli (bond) percolation threshold on the triangular lattice, $p_{\mathrm{c,tri}} \approx 0.35$ \cite{Sykes1964, Wierman1981} at a finite temperature. The phase diagram is structurally identical to the one on the cubic lattice \cite{Linsel2024}.

\begin{figure}[t]
\includegraphics[width=0.45\textwidth]{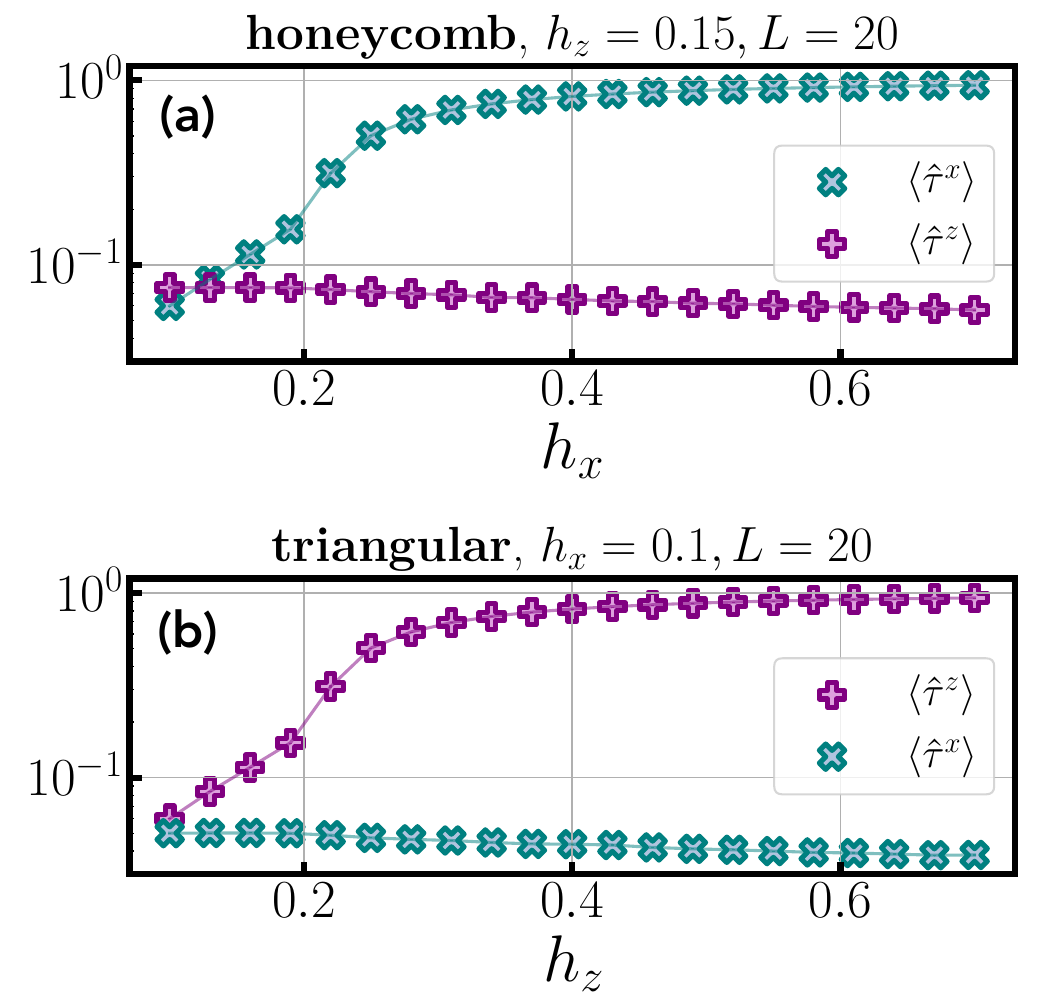}
\caption{\textbf{Magnetizations} in $x$- and $z$-direction of the extended toric code~(\ref{eq:eTC}). We apply continuous-time QMC at temperature $T=1/L$ for $L=20$. We calculate the $x$-magnetization $\langle\hat{\tau}^x\rangle$ and the $z$-magnetization $\langle\hat{\tau}^z\rangle$. (a) For large $h_x$ the system is magnetized in $x$-direction and completely paramagnetic in $z$-direction, here shown for the honeycomb lattice. (b) The opposite is true for large $h_z$, where the system is magnetized in $z$-direction and paramagnetic in $x$-direction, here shown for the triangular lattice.}
\label{fig:magnetization}
\end{figure}

\begin{figure*}[t]
\includegraphics[width=0.95\textwidth]{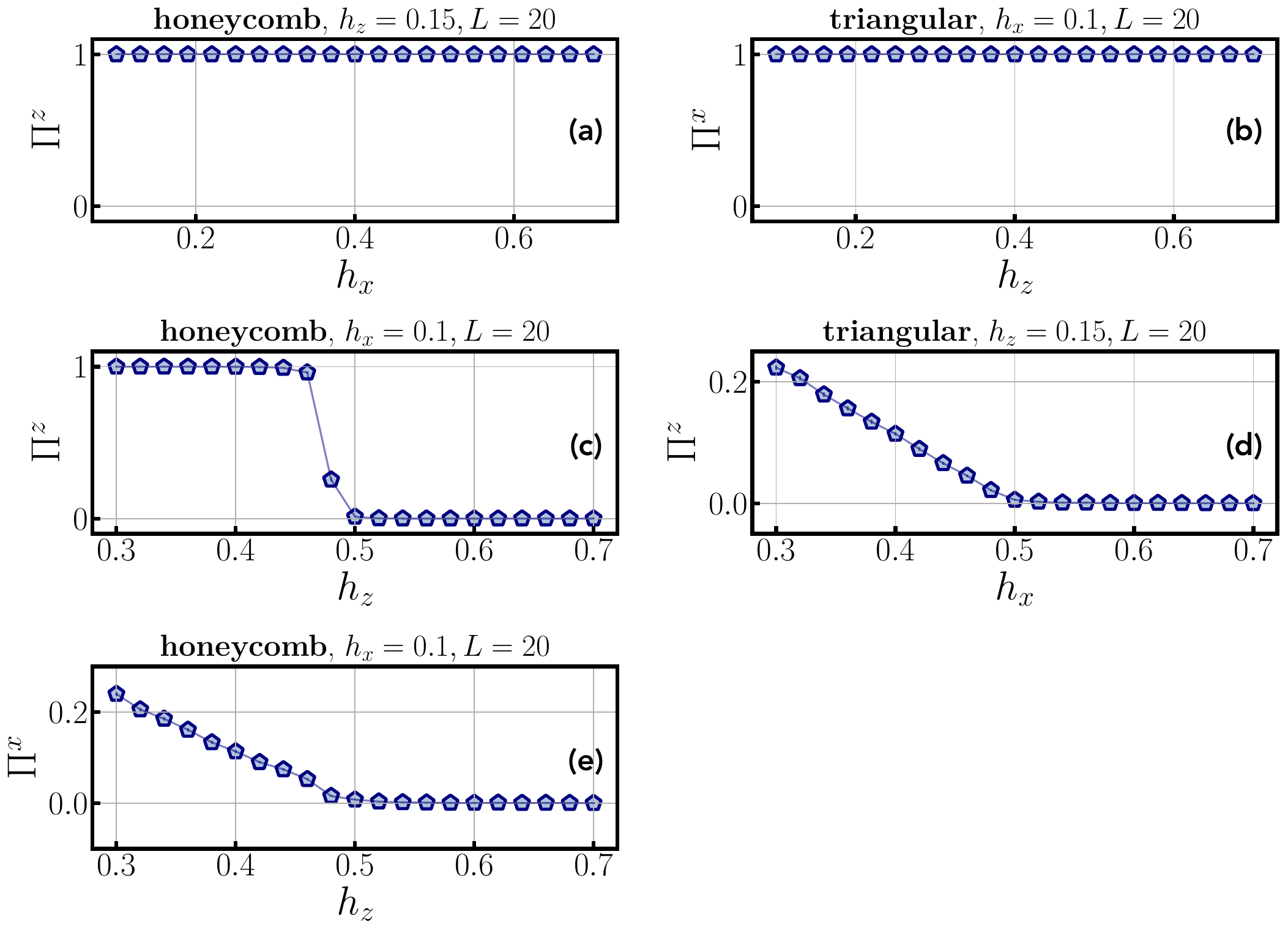}
\caption{\textbf{(Plaquette) percolation probabilities supporting the confinement phase diagram} Fig.~\ref{figConfinementPD} of the extended toric code~(\ref{eq:eTC}). We apply continuous-time QMC at temperature $T=1/L$ for $L=20$. We calculate the percolation probability $\Pi^x$ ($\hat{\tau}^x$-basis) and the plaquette percolation probability $\Pi^z$ ($\hat{\tau}^z$-basis), respectively. (a) For small $h_z$ the system is m-deconfined ($\Pi^z \neq 0$) on the honeycomb lattice even for large $h_x$, where the system is e-confined. Hence, e- and m-anyons are independently confined in this regime. (b) Similarly, for small $h_x$ the system is e-deconfined ($\Pi^x \neq 0$) on the triangular lattice even for large $h_z$, where the system is m-confined. (c-e) We show data for the phase boundaries for which no inset is included in Fig.~\ref{figConfinementPD}.}
\label{fig:details}
\end{figure*}

Our results can be directly related to the full quantum model studied in the main text. The topological ground states of the extended toric code~(\ref{eq:eTC}) feature percolating clusters since charges (associated with open ends $j$ of a cluster with $\prod_{l \in +_{\j}}\hat{\tau}_{l}^x = -1$) are gapped excitations and only appear in virtual pairs as a consequence of quantum fluctuations. In the Higgs phase ($h_z \gg 1$) on the other hand, charges condense and accumulate, leading to a finite density of free charges. Depending on the lattice geometry, this may prohibit percolation. On the triangular and cubic lattice (see Fig.~\ref{figConfinementPD}c,e) with larger coordination number, we observe $\Pi^x>0$ in the Higgs phase since the accumulation of charges does \textit{not} prohibit the formation of a percolating cluster, which is directly related to the fact that the percolating phase of the classical model~(\ref{eq:can_cl_ham}) is persistent for a finite matter density, see Fig.~\ref{fig:classical_pd}b. Conversely, on the square and honeycomb lattice with smaller coordination number we find $\Pi^x=0$ in the Higgs phase (extrapolated on the square lattice, see \cite{Linsel2024}), since any finite matter density prohibits percolation, as evident from Fig.~\ref{fig:classical_pd}a. Thus the classical finite-$T$ phase diagram is deeply connected to the behavior observed in the full quantum model and explains the qualitative differences between the lattices. Our work also connects to \cite{Hastings2014}, where a POP was applied at finite temperature in the context of quantum error correction.

%%%%%%%%%%%%%%%%%%%%%%%%%%%%%%%%%%%%%%%%%%%%%%%%%%%%%
\subsection{Additional details on the confinement phase diagram}
\label{app:details}
%%%%%%%%%%%%%%%%%%%%%%%%%%%%%%%%%%%%%%%%%%%%%%%%%%%%%

In Fig.~\ref{fig:magnetization}, we show the $x$-magnetization $\langle\hat{\tau}^x\rangle$ and the $z$-magnetization $\langle\hat{\tau}^z\rangle$ with respect to the fields $h_x$ and $h_z$. For high $h_x$ ($h_z$) the system is magnetized in the $x$-direction ($z$-direction) and paramagnetic in the $z$-direction ($x$-direction). Thus, for high $h_z$ the fields $\tau^x$ are completely independent and resemble Bernoulli percolation, where each bond is occupied with a probability $p=0.5$. Depending on the lattice geometry, this probability is below (honeycomb, $p_{\mathrm{c,hon}} \approx 0.65$ \cite{Sykes1964, Wierman1981}), at (square, $p_{\mathrm{c,sq}} = 1/2$ \cite{Sykes1964, Kesten1980}), or above (triangular, $p_{\mathrm{c,tri}} \approx 0.35$ \cite{Sykes1964, Wierman1981}; cubic,  $p_{\mathrm{c,cu}}=0.247(5)$ \cite{Sykes1964}) the percolation threshold. If $p_c<0.5$ for a given lattice, then the system is e-deconfined for small $h_x$ and $h_z \to \infty$. For plaquette percolation, the percolation threshold on the \textit{dual} lattice is decisive: the system is m-deconfined for small $h_z$ and $h_x \to \infty$ if $p_c>0.5$ for a given lattice. The special case $p_c=1/2$ for the square lattice is discussed in \cite{Linsel2024}.

In Fig.~\ref{fig:details}, we show additional details supporting the phase diagram presented in Fig.~\ref{figConfinementPD}. In Fig.~\ref{fig:details}a, we show the plaquette percolation probability $\Pi^z$ and tune $h_x$ across the e-confinement transition at $h_{x,\mathrm{c}} = 0.210(2)$. On both sides of the e-confinement transition and even for high fields $h_x$, the m-anyons are m-deconfined with $\Pi^z=1$. Thus, e- and m-anyons are independently confined. In Fig.~\ref{fig:details}b, we show the analogue on the triangular lattice. Here, e-anyons remain deconfined for high fields $h_z$ while m-anyons are simultaneously confined. In the topological phase at small fields $h_x, h_z$, both e- and m-anyons are deconfined. In Fig.~\ref{fig:details}c-e, we show exemplary plots of $\Pi^x, \Pi^z$ for the phase boundaries for which no inset is included in Fig.~\ref{figConfinementPD}.

\FloatBarrier

\end{document}